\documentclass[twocolumn,
aps,nofootinbib,showpacs,showkeys,preprint
tightenlines,preprintnumbers,] {revtex4}

\usepackage{epsf,epsfig,subfigure,graphicx,amsmath,amssymb}
\usepackage{color}
\newcommand{\dis}[1]{\begin{equation}\begin{split}#1\end{split}\end{equation}}

\newcommand{\OVER}[1]{\,\overline{\hskip -0.5mm #1}}

\newcommand{\etal}{{\it et al.}}

\newcommand{\gev}{\,\textrm{GeV}}

\newcommand{\eV}{\,\textrm{eV}}

\newcommand{\LCDM}{{$\Lambda$CDM cosmology}}

\def\phiq{{\phi_{\rm de}}}

\newcommand{\Z}{{\bf Z}}

\newcommand{\ie}{{\it i.e.~}}

\newcommand{\chiz}{{\chi^{(0)}}}
\newcommand{\chizc}{{\overline{\chi}^{\,(0)}}}

\newcommand{\Vde}{{$10^{-47}\,\gev^{\,4}$}}
\newcommand{\Vew}{{$v_{\rm ew}$}}

\newcommand{\UDE}{{U(1)$_{\rm de}$}}
\newcommand{\UPQ}{{U(1)$_{\rm PQ}$}}
\newcommand{\DEps}{{DEPS}}

\begin{document}
\draft

\title{\Large\bf Dark energy from approximate \UDE~ symmetry}  %Pursuit, Chase, Trace, Origin, Root, Essence

\author{Jihn E.  Kim$^{(a)}$ and Hans Peter Nilles$^{(b)}$ }
\affiliation
{
 $^{(a)}$
 Department of Physics, Kyung Hee University, 26 Gyungheedaero, Dongdaemun-Gu, Seoul 130-701, Republic of Korea,\\
 $^{(b)}$Bethe Center for Theoretical Physics and Physikalishes Institut der\\
   Universit\"at Bonn, Nussallee 12, 53115 Bonn, Germany
}

\begin{abstract}
The PLANCK observation strengthens the argument that the observed acceleration of the Universe is dominated by the invisible component of dark energy. We address how this extremely small DE density can be obtained in an ultraviolet complete theory. From two mass scales, the grand unification scale $M_G$ and the Higgs boson mass, we parametrize the scale of dark energy(DE). To naturally generate an extremely small DE term, we introduce an almost flat DE potential of a pseudo-Goldstone boson of an approximate global symmetry \UDE~ originating from some discrete symmetries allowed in an ultraviolet complete theory, as e.g. obtained in string theory constructions.
For the DE potential to be extremely shallow, the pseudo-Goldstone boson is required not to couple to the QCD anomaly. This fixes uniquely the
nonrenormalizable term generating the potential suppressed by $M_G^7$ in supergravity models.

\keywords{Dark energy, Quintessential pseudoscalar, Discrete symmetries}
\end{abstract}

\pacs{14.80.Va, 11.30.Er, 11.30.Fs, 95.36.+x}

\maketitle

%%%%%%%%%%%%%%%%%%%%%%%%%%%%%%%%%%%%%%%%%%%%%%%%%%%%
%%%%%%%%%%%%%%%%%%%%%%%%%%%%%%%%%%%%%%%%%%%%%%%%%%%%
\section{Introduction}
Recent observations \cite{Planck13} continue to support the existence of dark energy(DE), which is invisible to any method of detection except in Einstein's evolution equation \cite{Einstein17} of the Universe. The evidence for DE has been taken into account since the 1998 observation of the accelerated expansion of the Universe \cite{Perlmutter98}. The current DE under this hypothesis is extremely small compared to the fundamental energy density of gravity $M_P^4$, where $M_P\simeq 2.44\times 10^{18}\,\gev$. Even though there exist several attempts to interpret the accelerated expansion of the Universe, so far none of their parameters is explained from first principles. Therefore,  any good idea shedding light on this extremely small magnitude of
DE is welcome. In this paper, we suggest such a magnitude from a relation of the recently discovered Higgs boson mass \cite{Atlas12} compared to the Planck mass $M_P$. This relation leading to the small DE is drawn from the potential energy of an extremely light pseudo-Goldstone boson originating from a discrete symmetry principle in an ultraviolet complete theory
\cite{DiscreteString, DiscrGauge89, DiscreteR, KimMuSol13, KimPLB13, Kappl:2008ie, Nilles:2012cy}.
Such ultraviolet completions with abundant discrete symmetries can be found in the framework of explicit
(heterotic) string theory constructions in the sprit of
\cite{Kim:2006hw, Kim:2006hv, Lebedev:2006kn, Kim:2006zw, Lebedev:2007hv, Lebedev:2008un, Nilles:2008gq}.
They provide the quantum-gravity safe discrete symmetries as basis of an approximate global symmetry \UDE~ for generating the DE potential.

At present, DE is the dominant form of cosmic energy, constituting roughly 68\,\%, compared to 27\,\% CDM density \cite{Planck13}. The DE component drives the accelerated expansion in the \LCDM. But, at earlier times the DE was negligible compared to the CDM or radiation energies.
When CDM was the dominant component, the cosmological scale factor $a(t)$ grew as the $2/3$ power
law of the cosmic time, $\propto t^{2/3}$, and the current acceleration was driven  by comparing the data with the $2/3$ power law \cite{Perlmutter98}. It is the `coincidence puzzle' in particle cosmology why the DE density has overcome the CDM energy density quite recently in the cosmic time scale. In this paper, we will not attempt to attack this problem of coincidence puzzle, but the DE solution along our line of reasoning will be related to the axion CDM density.

%%%%%%%%%%%%%%%%%%%%%%%%%%%%%%%%%%%%%%%%%%%%%%%%%%%%%%%%%%%%
\begin{figure}[!t]
\begin{center}
\includegraphics[width=0.85\linewidth]{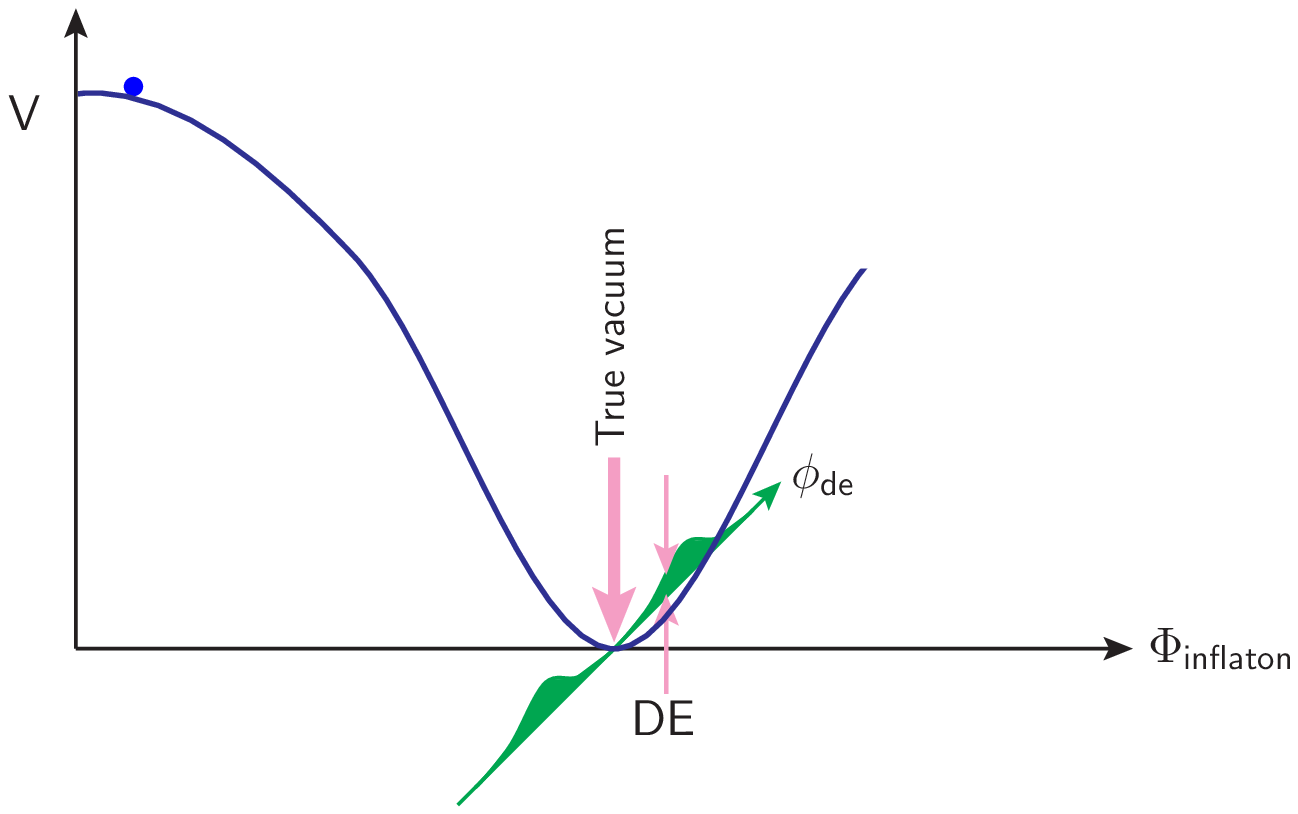} (a)
\includegraphics[width=0.85\linewidth]{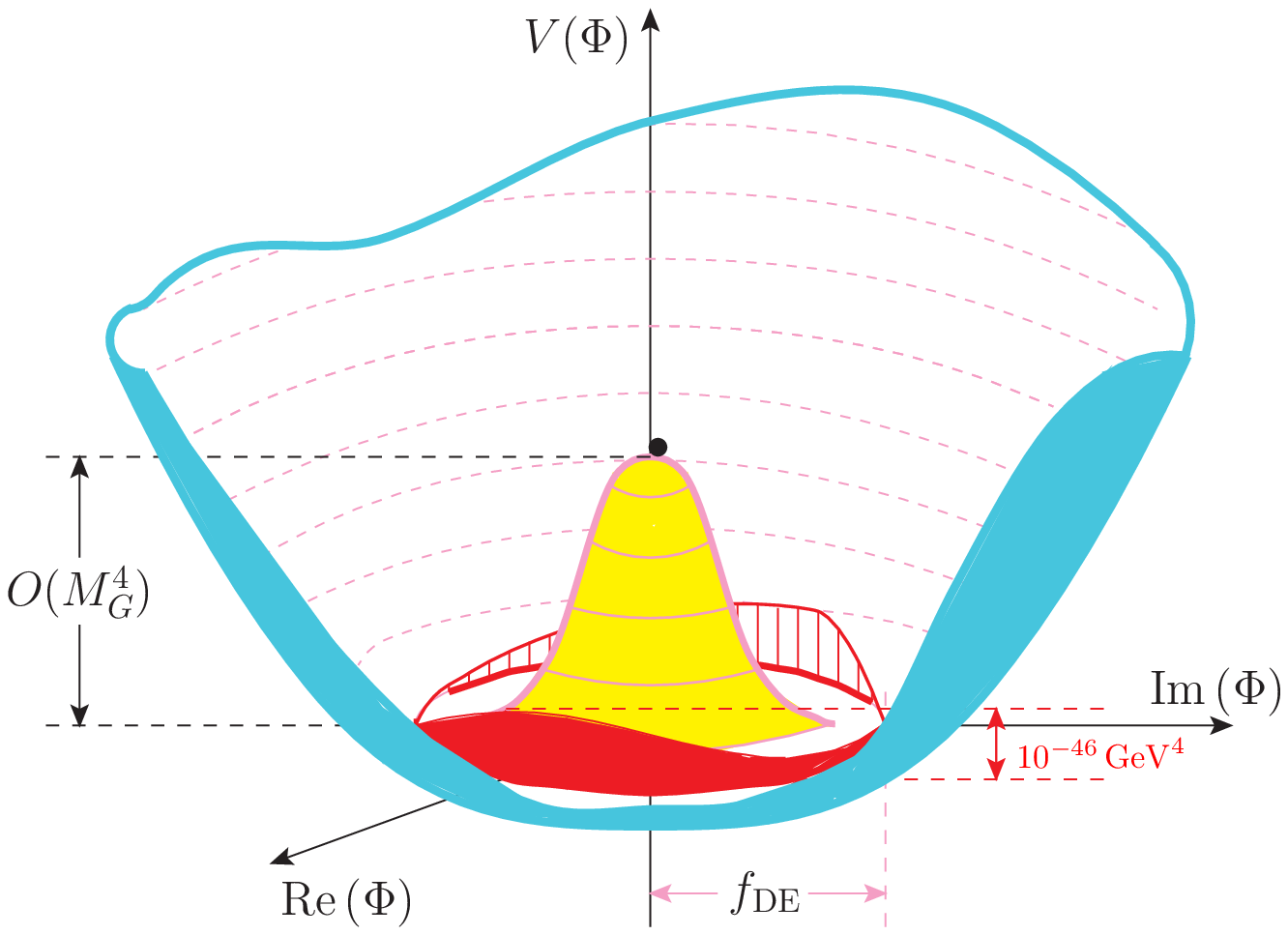} (b)
\end{center}
\caption{The dark energy  (DE)  of the Universe is represented by the green curve. (a) A possible inflaton potential is shown as well. The height of the green curve is exaggerated roughly by a factor of $10^{115}$.  (b) The red $\phiq$ potential with the height somewhat larger than \Vde~ is shown for $N_{DW}=2$, by closing the green of (a) as a circle. In this case, the `inflaton' can be $\Phi$, breaking \UDE, realizing a type of natural inflation \cite{NaturalInfl} in our scheme. } \label{fig:DEpotential}
\end{figure}
%%%%%%%%%%%%%%%%%%%%%%%%%%%%%%%%%%%%%%%%%%%%%%%%%%%%%%%%%%%%

In Fig. \ref{fig:DEpotential}\,(a), we present a picture for the potential energy densities  responsible for the time evolution of the Universe. The height of the green potential energy represents DE. The simplest form of DE is the cosmological constant (CC) in Einstein's equation \cite{Einstein17}. The theoretical CC problem has been to understand why the CC is zero at the vacuum where all equations of motion are satisfied. This vacuum is indicated by the thick lavender arrow in Fig. \ref{fig:DEpotential}(a). Even though we do not yet have a good theory for understanding the true vacuum with the vanishing CC, it is still meaningful to assume that the CC is zero at the true vacuum \cite{WeinbergRMP}. Then, the observed DE is evanescent, eventually converting into $\phiq$ oscillations as depicted by the green curves in Fig. \ref{fig:DEpotential}. The inflaton field $\Phi_{\rm inflaton}$ is responsible for the inflationary period in the very early Universe and the quintessential
pseudoscalar field $\phiq$ \cite{Frieman95, quintAx1, Chatzistavrakidis:2012bb}
is responsible for the recent accelerated expansion of the Universe by making the decay constant very large $\gtrsim M_P$.
Thus our discussion is based on the idea of a quintessential axion in the spririt of refs. \cite{Frieman95, quintAx1}.
We assume in this paper that the magnitude of DE is given by the height of the green shaded
area of the quintaxion potential in Fig. \ref{fig:DEpotential}\,(a).

This gives us an interesting connection to the idea of the mechanism  of an inflationary expenasion of the early universe.
We may identify $\Phi_{\rm inflaton}$ of Fig. \ref{fig:DEpotential}(a) as $\Phi$ of Fig. \ref{fig:DEpotential}(b) by
closing the green ``line'' of (a) to a circle. In this case,
a  scenario of the type of natural-inflation \cite{NaturalInfl} will appear,
which could lead to  {\it hilltop inflation} (at the hilltop of a Mexican hat potential).
Natural inflation is based on a cosine potential, and ours uses the quadratic potential as a leading term.
They are very similar because both of them give the common features for $\partial V/\partial v$ and
$\partial^2 V/\partial v^2$ near $v\approx 0$ where $v$ is the inflaton direction, \ie in our case $v=|\Phi|$.
They are very similar because natural inflation uses the shift symmetry of (a nonlinearly realized field)
$a_{\rm inflaton}\to a_{\rm inflaton}+{\rm constant}$ and the hilltop inflation uses the linear realization of \UDE.
If natural inflation is based on a confining force at a GUT scale, $a_{\rm inflaton}$ does not lead to small-field
inflation since the origin is the minimum as in the familiar QCD axion case. On the other hand, hilltop
inflation may be a small field inflation always due to the high temperature effects before the spontaneous
symmetry breaking of \UDE, as depicted as the bullet in Fig. \ref{fig:DEpotential}(b). Another difference is
that the inflaton in natural inflation is probably a pseudoscalar field while the inflaton in hilltop
inflation is a scalar field.

Apart from theses applications for QCD-axions and inflation,
the focus of this paper is on the dynamical origin of dark energy. Its main ingredient is an ultra light
quintessential axion with an axion decay constant of order of the string or GUT scale. The height of the
very shallow potential is the source of dark energy \cite{Frieman95, quintAx1, Chatzistavrakidis:2012bb}.
The scale of the potential is protected by an
approximate \UDE, that is derived from discrete symmetries in consistent string theory constructions
\cite{Kim:2006hw, Kim:2006hv, Lebedev:2006kn, Kim:2006zw, Lebedev:2007hv, Lebedev:2008un, Nilles:2008gq}.
These symmetries are of (discrete) gauge symmetry origin and will not be violated by gravitational quantum effects.

%%%%%%%%%%%%%%%%%%%%%%%%%%%%%%%%%%%%%%%%%%%%%%%%%%%%
%%%%%%%%%%%%%%%%%%%%%%%%%%%%%%%%%%%%%%%%%%%%%%%%%%%%
\section{Mass parameters from fundamental scalars}
The fundamental mass parameter at the Planck time is the reduced Planck mass $M_P\simeq 2.44\times 10^{18}\,\gev $ which
is related to Newton's gravitational constant $G= 1/8\pi M_P^2$.
in addition we will introduce the GUT mass $M_G\approx 0.01\, M_P$ for the parameter of suppression for higher dimensional operators. This is because our hypothetical fields are arising from some GUT multiplets, and if both the gravity contribution and the GUT contribution to the interactions are present then the GUT contribution dominates. The difficulty in any attempt to understand the magnitude of DE is its smallness compared to the energy scale at the Planck time, \ie $\sim M_P^4$. As a result of discovering the Higgs boson at $M_h\simeq\frac12$ \Vew~\cite{Atlas12},  a second fundamental mass parameter, \Vew~ originating from bosons, is now known to exist. \Vew~ is the vacuum expectation value(VEV) of the Higgs scalar at about $246\,\gev$. In terms of \Vew, all the known masses of the quarks and leptons are explained with suitable Yukawa couplings in the standard model(SM) of particle physics. Namely, \Vew~ provides all the masses of the SM, including $W^\pm$ and $Z^0$. Therefore, if DE can be calculated at all in terms of scalar VEVs, its simplest form is expressible in terms of two mass scales,
\dis{
M_G\approx 0.01\, M_P ,~ {\rm and}~ v_{\rm ew}.\label{eq:twoMasses}
}
The intermediate scale $M_{\rm int}\simeq\sqrt{v_{\rm ew}M_G}$ is parametrically dependent on $M_P$ and \Vew, and later the axion scale (about 100 times $M_{\rm int}$ \cite{InvAxionRev10}) will be used for it to include all the QCD anomaly couplings.
If $W/Z$ had obtained mass by the technicolor idea, the simple mass parameter to use at the TeV scale would have been the technicolor scale $\Lambda^3$, which will be very difficult to be implemented in our parameter fitting.

To compare our suggestion with some well-known suggestions for the acceleration of the Universe, we briefly comment how ours will be different from others in that the quantum-gravity safe discrete symmetries are employed or not. It is summarized in Table \ref{tab:Qnumb1}. We start with the so-called modified Newtonian dynamics(MOND) because it is most dramatically contrasted to our DE solution of the accelerating Universe.

%%%%%%%%%%%%%%%%%%%%%%%%%%%%%%%%%%%%%%%%%%%%%%%%%%%%%%%%%%%%%%%%%%%%%%%%%%%%%%%%%%%%%%
\begin{table}[b!]
\begin{center}
\begin{tabular}
{|c c|cc |}
\hline
{\rm Models}& References  &   ~Naturalness~   &
 ~Top--down scale~    \\[0.3em]
\hline\hline
MOND & \cite{MOND} & ~No~ & No\\[0.3em]
\hline
Anthropic & \cite{Weinberg87} & ~$???$~ & No\\[0.3em]
\hline
~Quint. PNGB & \cite{Frieman95} & ~Maybe~ & Maybe\\[0.3em]
\hline
Dilaton & \cite{Wetterich88} & ~Maybe~ & No\\[0.3em]
\hline
Quint. scalar & \cite{Steinhardt99, Albrecht00} & ~No~ & Maybe\\[0.3em]
\hline
Quint. axion & \cite{quintAx1} & ~Maybe~ & Maybe\\[0.3em]
\hline
 ~\DEps~ &  &  ~Yes~ &  Yes\\[0.3em]
\hline
\end{tabular}
\end{center}\caption{Comparison of our DEPS with other explanations for the SNIa data.
}\label{tab:Qnumb1}
\end{table}
%%%%%%%%%%%%%%%%%%%%%%%%%%%%%%%%%%%%%%%%%%%%%%%%%%%%%%%%%%%%%%%%%%%%%%%%%%%%%%%%%%%%

One obvious attempt to explain the recent acceleration is the MOND. In MOND, Newton's law is changed by introducing an acceleration parameter $a_0\simeq 1.2\times 10^{-8}\,{\rm cm\,s^{-2}}$ at the cosmic scale where the measured acceleration was reported. With this, the rotation curves of most galaxies can be explained without the need for CDM \cite{MOND}. But, MOND fails to explain DM at the cluster scale of galaxies and more importantly the primordial production of light elements, $^2$H, $^3$He, and $^7$Li. Then,  MOND also needs the CDM component for nucleosynthesis. Comparing MOND and \LCDM~ in deriving their input parameters from scalar masses in an ultraviolet completed theory, it may be more difficult to obtain $a_0$ in a MOND \cite{Sarkar13} than to obtain a reasonable DE scale in a \LCDM ~as shown in this paper.

Since the DE amount is only about 2.5 times that of CDM, the coincidence, ``Why is the amount of DE comparable to the amount of matter today?", is intriguing and attempted to be understood by changing the equation of state, most probably via the potential energy of a scalar field \cite{Steinhardt99, Albrecht00, Copeland00, Copeland06}. In this regard, the (nonlinearly realized) dilaton was suggested and the dilatonic symmetry was assumed with the spontaneous symmetry breaking scale at $\sim M_P$ \cite{Wetterich88, Ratra88}. The explicit breaking scale of the dilatonic symmetry is via the dimensional transmutation of asymptotically
free theories, but the dimensional transmutation scale is not a mass parameter of scalar fields. \Vew~ determined
from mass parameters of scalar fields cannot serve as an explicit symmetry breaking of the dilatonic symmetry because
it is another VEV breaking the dilatonic symmetry. Another feature of dilaton toward a DE solution is  changing Newton's
constant, which is unsatisfactory at the moment. In addition, we do not find any discussion on a basic discrete symmetry
which we adopt here.

There has been a tremendous effort to understand the DE scale in the \LCDM,
not changing Newtonian dynamics. The simplest account of DE in this direction is the CC itself, but
there is a theoretical difficulty to consider an extremely small CC as commented above that there is not yet a self-tuning solution  towards a vanishing CC \cite{Foerste13, Selftune, SelfHorndeski}.
So, if the CC itself is considered as the observed DE, the anthropic bound  $\rho_{\rm DE} < 550\,\rho_{\rm CDM}\simeq (5\times 10^{-3}\,\eV)^4$ \cite{Weinberg87} is the most plausible argument.

In Table \ref{tab:Qnumb1}, naturalness in the second column is judged from the possibility of obtaining it from a
symmetry principle, in particular from a discrete symmetry principle. We shall explore the possibility to
obtain all relevant mass parameters from two mass scales, $M_G\approx 0.01\,M_P$ and \Vew.

%%%%%%%%%%%%%%%%%%%%%%%%%%%%%%%%%%%%%%%%%%%%%%%%%%%%%%%%%%%%%%%%%%%%%%%%%%%%%%%%%
%%%%%%%%%%%%%%%%%%%%%%%%%%%%%%%%%%%%%%%%%%%%%%%%%%%%%%%%%%%%%%%%%%%%%%%%%%%%%%%%%
\section{\UDE~ and Goldstone boson}
Thus, we attempt to introduce a flat potential first and then raise it by a tiny amount. For this, it is necessary to introduce a massless particle in the first step.
The most plausible theory obtaining an extremely light particle is to trigger spontaneous symmetry breaking of a global U(1) symmetry, leading to a massless Goldstone boson with parity $-1$ \cite{Goldstone61}. It appears in the imaginary exponent of the VEV generating complex scalar field, $(v+\rho)e^{i\theta}$. The VEV, $f_{\rm DE}$ of Fig. \ref{fig:DEpotential}(b), of this complex scalar is taken at the Planck scale and the explicit breaking term of the global U(1) symmetry, making it a pseudo-Goldstone boson, is at the observed DE scale, somewhat bigger than \Vde~ \cite{Frieman95}. Even though the height of Fig. \ref{fig:DEpotential}\,(b) is of order $M_G^4$, the decay constant $f_{\rm DE}$ can be trans-Planckian via a small quartic coupling $\lambda^2(\lesssim 10^{-4})$ of the \UDE~ breaking field $\Phi$,
\dis{
V= \frac{\lambda^2}{4} (\Phi^*\Phi)^2-M_G^2\Phi^* \Phi+{\rm constant}.
}
Here, we do not have the difficulty encountered for the case of $B_{MN}$ in raising their decay constants to $M_P$.
String theory has the gravity multiplet in 10-dimensions $(M,N=0,1,2,\cdots,9)$: the symmetric tensor field $g_{MN}$ which contains graviton, the dilaton, and the antisymmetric tensor field $B_{MN}$. When six of ten dimensions are compactified,  $B_{MN}$ leads to pseudoscalar fields, the model-independent(MI) pseudoscalar for $\{M,N\}=\{0,1,2,3\}$ and model-dependent(MD) pseudoscalar for $\{M,N\}=\{4,\cdots,9\}$.  These were considered before for the DE called `quintessential axion' \cite{quintAx1}.
It is known that the quintessential axion has a difficulty for making its potential extremely flat unless massless
quarks are introduced \cite{QuintAx2}. Probably, the MD axions obtain non-negligible superpotential terms \cite{WenWitten},
which is the reason we usually neglect these at low energy. For the MI axion, it can become the phase of a \UPQ~
transformation below the string scale if the compactification leads to the anomalous U(1). The very light QCD axion
from string theory is usually based on this scheme \cite{Kim88}.
Because the decay constants of $B_{MN}$ turn out to be somewhat reduced from $M_P$ \cite{Svrcek}, there is a difficulty
obtaining a trans-Planckian decay constants for $B_{MN}$, and hence 2-flation \cite{Peloso04} and N-flation \cite{Nflation}
have been considered. Because $\phiq$ in our case is not coming from $B_{MN}$ but rather from matter fields
of the  E$_8\times$E$_8'$ representations of the heterotic string in the models considered here
\cite{Kim:2006hw, Kim:2006hv, Lebedev:2006kn, Kim:2006zw, Lebedev:2007hv, Lebedev:2008un, Nilles:2008gq},
we can easily obtain a trans-Planckian $f_{\rm DE}$ by an O($10^{-2}$) coupling.
Thus, the spontaneous symmetry breaking scale of \UDE~ is
\dis{
f_{\rm DE}= \sqrt{M_P^2+M^2_{\rm int}}\gtrsim M_P\label{eq:fde}
}
can be of trans-Planckian as depicted in Fig. \ref{fig:DEpotential}(b). In Eq. (\ref{eq:fde}), $M_{\rm int}$ is assumed to break  also the \UDE~ symmetry. Since the height of the GUT scale breaking $M_G^4$ of \UDE~ is smaller than $M_P^4$ by a factor of $\sim 10^{-8}$, there is not much gravity interference of the $f_{\rm DE}$ determination.
A generic global symmetry is generically spoiled by gravitational effects \cite{KimPLB13, Giddings87, BarrPQ92},
but here we use (discrete) gauge symmetries from string theory that do not suffer from this problem.

In this paper we introduce a `$\underline{\rm D}$ark $\underline{\rm E}$nergy $\underline{\rm p}$eudo$\underline{\rm s}$calar boson'(DEPS) which does not couple to the QCD (and hidden-sector non-Abelian anomalies, if present). The corresponding approximate global symmetry is called \UDE, and the DEPS $\phiq$ is the corresponding pseudo-Goldstone boson. The first step toward obtaining the DE scale of the Universe is to have an exactly massless Goldstone boson $\phiq^{(\rm def)}$, with superscript $^{(\rm def)}$ meaning the massless Goldstone boson \cite{Goldstone61} from the \UDE-defining terms.

To relate the DEPS to \Vew, it is necessary to couple it to the Higgs fields $H_uH_d$, which implies that \UDE~ has a
QCD anomaly because $H_u$ and $H_d$ couple to quarks. Therefore, to have a QCD-anomaly free \UDE, it is necessary to
introduce two U(1) global symmetries, \UDE~ and \UPQ~ in our case so that one  anomaly-free combination results.  The
anomaly-free combination is our DEPS. The remaining one is the QCD axion. So, in our discussion the appearance of the
QCD axion is inescapable.
In the next step we break the anomaly free symmetry slightly
to obtain a pseudo-Goldstone boson $\phiq$ that feebly contributes to the vacuum energy via the
terms in the red part of Fig. \ref{fig:DEpotential}(b).

%%%%%%%%%%%%%%%%%%%%%%%%%%%%%%%%%%%%%%%%%%%%%%%%%%%%%%%%%%%%%%%%%%%%%%%%%%%%%%%%%
%%%%%%%%%%%%%%%%%%%%%%%%%%%%%%%%%%%%%%%%%%%%%%%%%%%%%%%%%%%%%%%%%%%%%%%%%%%%%%%%%
\section{Exact discrete symmetry and pseudo-Goldstone boson}

%%%%%%%%%%%%%%%%%%%%%%%%%%%%%%%%%%%%%%%%%%%%%%%%%%%%%%%%%%%%
\begin{figure}[!t]
\begin{center}
\includegraphics[width=0.6\linewidth]{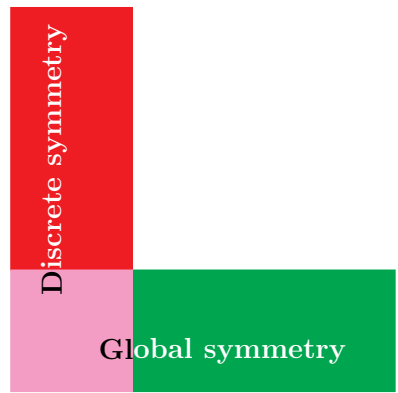}
\end{center}
\caption{A sketch for the terms satisfying discrete (red) and global (green) symmetries. The lavender part is the common intersection. } \label{fig:GlobalRegion}
\end{figure}
%%%%%%%%%%%%%%%%%%%%%%%%%%%%%%%%%%%%%%%%%%%%%%%%%%%%%%%%%%%%

We propose to use suitable  discrete symmetries toward obtaining our desired approximate global symmetry \UDE.
Even before the 1998 discovery of the accelerating Universe, discrete symmetries were considered for obtaining some approximate global symmetry \cite{Garretson93},
but being before the 1998 discovery, only a general setup could be given. Ours is the first serious one using a discrete symmetry towards obtaining a (transient) non-vanishing CC, and along the way  we have already commented on a new idea of hilltop inflation.
Of course, the hypothetical discrete symmetry must satisfy the discrete gauge symmetry rule \cite{DiscrGauge89},
as it happens in the string model constructions considered here.

Gauge symmetries are not spoiled by gravitational interactions. If a discrete symmetry results from a subgroup of gauge symmetries of string compactifications, gravity does not spoil the discrete symmetry \cite{KimPLB13}. One can consider a series of interaction terms allowed by the discrete symmetry. This infinite tower of terms, not spoiled by gravity, is shown as the vertical red bar in Fig. \ref{fig:GlobalRegion}. If one considers a few lowest order terms of the red column, we can find an accidental global symmetry. Using this global symmetry, one can consider an infinite series of terms as marked in the horizontal green bar in Fig. \ref{fig:GlobalRegion}. The terms shown in the lavender part of the vertical column, containing the \UDE~ defining terms, satisfy both the discrete and global symmetry transformations. But, the horizontal green bar terms outside the lavender are spoiled by gravity and hence we will not consider them. The vertical red bar terms outside the lavender are not spoiled by gravity, but break the global symmetry. This red part is the source for \Vde, making the Goldstone boson the pseudo-Goldstone boson and generating the DE scale.

We identify the \UPQ~ as the anomalous U(1) of string theory for the QCD axion \cite{Kim88} which is spontaneously broken at the intermediate scale. Since we will require the \UDE~ not carrying the anomaly including the anomalous U(1), the spontaneous symmetry breaking scale is generically around the Planck scale, $\gtrsim M_P$, as commented above. This is the picture by which we introduce the height of the $\phiq$ potential.

%%%%%%%%%%%%%%%%%%%%%%%%%%%%%%%%%%%%%%%%%%%%%%%%%%%
%%%%%%%%%%%%%%%%%%%%%%%%%%%%%%%%%%%%%%%%%%%%%%%%%%%
\section{Fitting DE scale to $\phiq$ potential}
Below, for an explicit presentation we work in the so-called $N=1$ supersymmetric (SUSY) extension of the SM \cite{Nilles84}. The DE scale is expressed by the VEVs of $H_u$ and $H_d$, \ie $v_u$ and $v_d$ which are of order \Vew. The intermediate scale is defined as
\dis{
M_{\rm int}\simeq\sqrt{v_{\rm ew}\, M_G}\simeq 2.5\times 10^{9}\,\gev
}
where we used $M_G\approx 2.5\times 10^{16}\,\gev$. The DEPS $\phiq$ originates from  the complex scalars $\chiz$ and $\chizc$ whose VEVs are comparable to the axion decay constant of order $10^{11-12}\,\gev$ related to the QCD axion.
So, $\langle\chi^0\rangle/M_{\rm int}$ is of order $10^2$.

%%%%%%%%%%%%%%%%%%%%%%%%%%%%%%%%%%%%%%%%%%%%%%%%%%%%%%%%%%%%
\begin{figure}[t!]
\begin{center}
\includegraphics[width=0.85\linewidth]{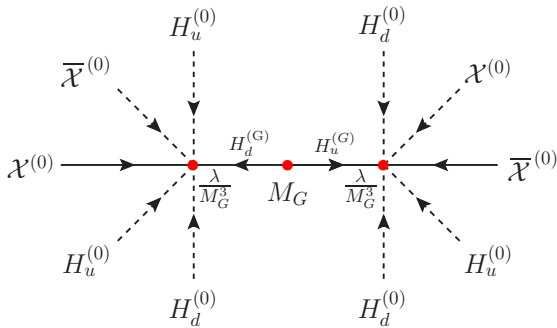}
\end{center}
\caption{The leading \UDE~ violating diagram.
} \label{fig:TwoDEpsino}
\end{figure}
%%%%%%%%%%%%%%%%%%%%%%%%%%%%%%%%%%%%%%%%%%%%%%%%%%%%%%%%%%%%

Including the soft supersymmetry breaking $A$-term
containing one factor of $m_{3/2}$ \cite{Nilles:1982dy, Nilles84}, we need an odd number of $M_G$ suppression factor so that the resulting potential is split into two groups with the equal number of scalar fields.
So, \Vde~ is parametrically expressible in terms of \Vew~ and $M_G$ as
\dis{
10^{8}\, \frac{ m_{3/2} v_{\rm ew}^{8}}{M_G^5} \sim 10^{-47}\,\gev^4 \label{eq:chiToVew}
}
where $m_{3/2}$ is the TeV scale gravitino mass. For $m_{3/2}v_{\rm ew}^3\approx 10^8\,\gev^4$ and $v_{\rm ew}/M_G\simeq 10^{-14}$, this height is roughly $10^{-44}\,\gev^4$. There is some unknown factor in $m_{3/2}v_{\rm ew}^3$ and hence the $M_G^{-5}$ suppression is considered as an adequate one. With $M_G^{-3}$ suppression, the potential is too large compared to \Vde, and we would have gone through  DE domination much earlier, which would not correspond to the current universe. With $M_G^{-7}$ suppression, the potential is too shallow to have any effect on the recent history of the Universe. In any case, there are three relevant suppression factors, $M_G^{-1, -3, -5}$. Out of these, $M_G^{-1 }$ is responsible for the $\mu$ term and the axion \cite{KimNilles84, KimMuSol13}.

In fact, within a scheme based on supergravity, we can consider effective superpotential terms ordered in powers of $1/M_G$,
\dis{
W=W^{(3)} + \sum_{i=4}^{\infty} \left( \frac{c_i}{M_G^{i-3}}\right)\, W^{(i)}.
}
Here, $ W^{(4)}$ defines the PQ symmetry which is explicitly broken by the QCD anomaly. Our definition of the PQ symmetry is {\it \`a la} Ref. \cite{KimNilles84}, $H_uH_dX\OVER{X}/M_G$ with singlet scalar fields $X$ and $\OVER{X}$.
$ W^{(4)}$ also contains the Weinberg operator $H_u^2\ell\ell/M_G$ where $\ell$ is the lepton doublet in the SM.
The \UDE~ symmetry is given by $W^{(6)}$.
To forbid \UDE~ symmetric terms in $W^{(4,5)}$, we typically need dicrete symmetries of large order $N$ as e.g. $\Z_{NR}$.
A realization of the Weinberg operator in terms of renormalizable terms is the seesaw model \cite{Seesaw}. A realization of the Kim-Nilles operator in terms of renormalizable terms is the ``invisible'' axion model \cite{VeryLightAxion}. For the roles of $W^{(4,5,6)}$ toward the cosmology of pseudo-Goldstone bosons, $W^{(4)}$ defines the QCD axion.  $W^{(5)}$, if present, would not have led to our Universe because of too much CDM without DE at present.
$W^{(7)}$, even if present, would not have affected our Universe so far.
Therefore, only the cases  $W^{(5)}$ and $W^{(6)}$ are relevant for our discussion of DE.
Anthropic arguments might be used to select from the landscape of discrete symmetries of the underlying models,
resulting in
pseudo-Goldstone bosons
from  $W^{(5)}$ and $W^{(6)}$. Being discrete, it is quite possible that an O($0.1$) fraction of the allowed models forbids  $W^{(5)}$ and chooses $W^{(6)}$. $W^{(10)}$ contains the term suppressed by $M_G^7$ for Fig. \ref{fig:DEpotential}\,(b).

\begin{figure}[!t]
\begin{center}
\includegraphics[width=0.6\linewidth]{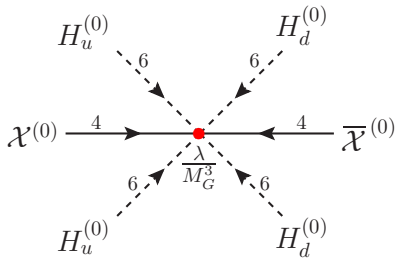}
\end{center}
\caption{The diagram as defined by the global symmetries. The numbers are the quantum numbers of U(1)$_{10R}$ gauge symmetry which become $\Z_{10R}$ charges mudulo 10.} \label{fig:GlobalDef}
\end{figure}

In the following we will sketch the qualitative picture of our mechanism. We need discrete symmetries of large order.
The models involve a large number of fields and a complete definition of the model is beyond the scope of this paper.
To see the details of the model, please consult ref. \cite{KimMore}, where all symmetries are displayed. The basic
picture considers the coupling to $H_uH_d$,  as in the generation of the axion scale in \cite{KimNilles84} through
$H_uH_dX\OVER{X}/M_G$, but here we have to consider larger symmetries and higher powers of the superpotential.
For this purpose we introduce new fields $\chi^0$ and $\OVER{\chi}^{\,0}$ (not to be confused with the $X$ and $\OVER{X}$
discussed in \cite{KimNilles84}). If in the superpotential
$j$ factors of $H_uH_d$ (\ie \Vew$^2$) are replaced by $2j$ factors of $\chi^0\OVER{\chi}^{\,0}$, with $\langle\chi^0\rangle/M_{\rm int}\sim 10^2$, we obtain an enhancement factor of $10^{4j}$.
Thus, two powers of $v_{\rm ew}$ (\ie $j=2$) in Eq. (\ref{eq:chiToVew}) are traded for the intermediate scale VEVs, $v_{\rm ew}^2\sim 10^{-8}\langle\chiz\rangle^2 \langle\chizc\rangle^2/M_G^2$, to obtain the height of the DEPS potential as
\dis{
V\sim m_{3/2}\frac{v_{\rm ew}^{6} \langle\chiz\rangle^2 \langle\chizc\rangle^2}{M_G^7}.
}
Anyway, relating $\chiz\chizc$ to $v_{\rm ew}$ is needed to make DEPS not couple to the QCD anomaly.
This height represents the breaking of \UDE~ and is generated by the red part terms of Fig. \ref{fig:GlobalRegion}. Let this be composed of $2n$ external lines (2 fermion lines and $(2n-2)$ boson lines) in SUSY with dimension $(2n+1)$ since there are two external fermion lines. So, it has the mass suppression factor $(1/M_G^{2n-3})$, determining $n=5$, \ie 2 fermion lines and 8 boson lines. Fig. \ref{fig:TwoDEpsino} shows a typical $A$-term realization of the potential in supergravity with the heavy internal line with external lines of three $H_u^{(0)}$, three $H_d^{(0)}$, two $\chiz$, and two $\chizc$. Fig. \ref{fig:TwoDEpsino} breaks the \UDE~ symmetry if this diagram is obtained by connecting two \UDE~ defining diagrams with one common fermion line connecting them. For example, the \UDE~ defining diagram consists of two fermion lines and four boson lines as shown in Fig. \ref{fig:GlobalDef}.  The quantum numbers are those of U(1)$_{10R}$ gauge symmetry which become the $\Z_{10R}$ charges mudulo 10. This is obtained from the left/right part of Fig. \ref{fig:TwoDEpsino}.  So, the \UDE~ defining diagram of Fig. \ref{fig:GlobalDef} is
\dis{
W\sim \frac{1}{M_G^3}\left(H_u^{(0)} H_d^{(0)}\right)^2\chiz\chizc. \label{eq:DSdef}
}
$H_{u,d}^{(G)}$ of Fig. \ref{fig:TwoDEpsino}  carry the same discrete quantum numbers of
$H_{u,d}^{(0)}$ of Fig. \ref{fig:GlobalDef}, and  Fig. \ref{fig:TwoDEpsino} breaks the \UDE~ global symmetry.
In our use of discrete symmetries as the origin of DEPS, Eq. (\ref{eq:DSdef}) is chosen as a definition of the approximate global symmetry from a few interaction terms in the lavender part of Fig. \ref{fig:GlobalRegion}. The discrete $\Z_{10R}$ quantum numbers are $\Z_{10R}(H_u^{(0)}, H_d^{(0)})=6$  and $\Z_{10R}(\chiz, \chizc)=4$ (see ref. \cite{KimMore}).  The term $M_G H_u^{(0)} H_d^{(0)}$ can be forbidden by a permutation discrete symmetry \cite{KimMuSol13}.
Then, the pseudo-Goldstone boson mass is read from Fig. \ref{fig:TwoDEpsino}.

%%%%%%%%%%%%%%%%%%%%%%%%%%%%%%%%%%%%%%%%%%%%%%%%%%%%%%%%%%%%
%%%%%%%%%%%%%%%%%%%%%%%%%%%%%%%%%%%%%%%%%%%%%%%%%%%%%%%%%%%%

So, to have both the QCD axion and the quintessential \DEps, we can introduce $X$ and $\OVER{X}$ type fields together with  $\chi$ and $\overline{\chi}$ type fields, and consider two approximate U(1) global symmetries. The U(1)$_{\rm PQ}$ is designed to carry all the color anomaly U(1)$_{\rm PQ}$--SU(3)$_c$--SU(3)$_c$,  while \UDE~ does not have the color anomaly. This is achieved by introducing heavy quarks $Q$ and  $\OVER{Q}$ \cite{KimMore}.

%%%%%%%%%%%%%%%%%%%%%%%%%%%%%%%%%%%%%%%%%%%%%%%%%%%
%%%%%%%%%%%%%%%%%%%%%%%%%%%%%%%%%%%%%%%%%%%%%%%%%%%
\section{Conclusion}

Unlike the other ideas presented in Table \ref{tab:Qnumb1}, the DEPS idea can have a naturalness origin from a
discrete symmetry principle and the height of the potential as low as $10^{-46}\,\gev^4$ can be obtained in terms
of parameters of Eq. (\ref{eq:twoMasses}): $M_G\approx 0.01\,M_P$ and \Vew.

To obtain an extremely small DE term, one needs an extremely light bosonic particle protected by a dicsrete symmetry
of high order. If its mass is
tiny, its potential energy near the origin is almost zero. So, as the starting point towards obtaining the extremely small value for DE,
we have taken the road to start with a massless Goldstone boson \cite{Goldstone61}.
For ordinary global symmetries, wormholes and black holes destroying
global charges have been the stumbling block for obtaining such a Goldstone boson.
Our strategy using discrete (gauge) symmetries from ultraviolet consistent string theory constructions
removes this stumbling blocks and opens the road toward a realistic origin of DE
with the help of a very very light pseudo-Goldstone boson. The scheme includes a QCD axion (to remove the QCD anomaly
from the quintessential axion) and can be extended to a model of natural or hilltop inflation.

\vskip2cm

%%%%%%%%%%%%%%%%%%%%%%%%%%%%%%%%%%%%%%%%%%%%%%%%%%%%%%%%%%%%%%%%%%%%%%%
\section*{Acknowledgments}

J.E.K. is supported in part by the National Research Foundation (NRF) grant funded by the Korean Government (MEST)
(No. 2005-0093841). H.P.N. is partially supported by SFB-Transregio TR33 ``The Dark Universe"
(Deutsche Forschungsgemeinschaft) and the European Union 7th network program ``Unification in the LHC era" (PITN-GA-2009-
237920).

%\end{document}
%%%%%%%%%%%%%%%%%%%%%%%%%%%%%%%%%%%%%%%%%%%%%%%%%%%%%%%%%%%%%%%%%%%%%%%%%%
%%%%%%%%%%%%%%%%%%%%%%%%%%%%%%%%%%%%%%%%%%%%%%%%%%%%%%%%%%%%%%%%%%%%%%%%%%
%%%%%%%%%%%%%%%%%%%%%%%%%%%%%%%%%%%%%%%%%%%%%%%%%%%%%%%%%%%%%%%%%%%%%%%%%%%%%%%%%%

%\newpage

\end{document}